\documentclass{egpubl}
\usepackage{pg2025}
\SpecialIssuePaper         
\CGFStandardLicense

\usepackage[T1]{fontenc}
\usepackage{dfadobe}  

\usepackage{cite}  
\BibtexOrBiblatex
\electronicVersion
\PrintedOrElectronic
\ifpdf \usepackage[pdftex]{graphicx} \pdfcompresslevel=9
\else \usepackage[dvips]{graphicx} \fi

\usepackage{egweblnk} 
\usepackage{enumitem}
\usepackage{amsmath}
\usepackage{amsfonts}
\usepackage{multirow}
\usepackage{bigdelim}
\usepackage{color}
\usepackage{booktabs}
\usepackage{ifthen}
\usepackage{hyperref}
\usepackage{subcaption}
\usepackage{bbold}
\usepackage{gensymb}
\usepackage{xspace}
\usepackage[noabbrev,capitalise,nameinlink]{cleveref}
\creflabelformat{equation}{#2\textup{#1}#3}%
\usepackage{xcolor, pgf}
\usepackage{color, colortbl}
\usepackage[normalem]{ulem} 
\usepackage{xr}
\usepackage{placeins}
\usepackage{soul}
\usepackage[many]{tcolorbox}
\usepackage{tikz}
\newcommand{\redline}{\raisebox{1.5pt}{\tikz{\draw[-,red,dashed,line width = 1pt](0,0) -- (6mm,0);}}}
\newcommand{\blackline}{\raisebox{1.5pt}{\tikz{\draw[-,black,dashed,line width = 1pt](0,0) -- (6mm,0);}}}
\usepackage{adjustbox}
\usepackage{algorithm}
\usepackage{algorithmicx}
\usepackage{algpseudocode}
\algnewcommand\algorithmicinput{\textbf{Input:}}
\algnewcommand\INPUT{\item[\algorithmicinput]}
\algnewcommand\algorithmicoutput{\textbf{Output:}}
\algnewcommand\OUTPUT{\item[\algorithmicoutput]}
\algnewcommand\algorithmicforeach{\textbf{for each}}
\usepackage{diagbox}
\usepackage{slashbox}
\usepackage[normalem]{ulem}

\algdef{S}[FOR]{ForEach}[1]{\algorithmicforeach\ #1\ \algorithmicdo}
\algrenewcommand{\alglinenumber}[1]{\color{red!80!blue}\footnotesize#1:}

\algnewcommand\Func[2]{\textcolor{green}{#1}\textcolor{green}{(#2)}}
\algnewcommand\Insert[2]{Insert {#1} to #2.}
\algnewcommand\Input[1]{\State \textbf{Input: } #1}
\algnewcommand\Output[1]{\State \textbf{Output: } #1}

\usepackage{eqparbox}
\newdimen{\algindent}
\algnewcommand\LeftComment[2]{%
\hspace{#1\algindent}$\triangleright$ \eqparbox{COMMENT}{\textcolor{blue}{#2}} \hfill %
}

\newlength\myboxwidth
\definecolor{gray}{rgb}{0.5,0.5,0.5}
\definecolor{green}{rgb}{0, 0.6, 0}
\definecolor{orange}{rgb}{1, 0.5, 0}
\definecolor{mahogany}{rgb}{0.75, 0.25, 0.0}
\definecolor{purple}{rgb}{0.6, 0, 0.6}
\definecolor{darkgreen}{rgb}{0, 0.3, 0}
\definecolor{orange}{rgb}{1, 0.5, 0.}
\definecolor{purpleD}{rgb}{.8941, .8, .9043}
\definecolor{greenD}{rgb}{.7145, .9249, .7999}
\definecolor{lightblue}{rgb}{0.52, 0.75,0.91}
\definecolor{secondblue}{rgb}{0.52, 0.78,0.70}
\definecolor{softgreen}{rgb}{0.66,0.87,0.74}
\definecolor{softred}{rgb}{0.96,0.71,0.69}

\newcommand{\methodName}{LayoutRectifier\xspace}

\newcommand{\bestcell}[1]{\cellcolor{lightblue!50}#1}
\newcommand{\seccell}[1]{\cellcolor{secondblue!50}#1}

\colorlet{soullightblue}{lightblue!50}
\newcommand{\besthint}[1]{\sethlcolor{soullightblue}\hl{#1}}
\colorlet{soulsecondblue}{secondblue!50}
\newcommand{\sechint}[1]{\sethlcolor{soulsecondblue}\hl{#1}}
\colorlet{soullightyellow}{yellow!50}
\newcommand{\sighl}[1]{#1}
\newcommand{\pghl}[1]{#1}
\newcommand{\rvhl}[1]{#1}

\newcolumntype{a}{>{\columncolor{lightblue!50}}c}

\newcommand{\ignore}[1]{}
\newcommand{\none}[1]{}
\newcommand{\com}[1]{}
\newcommand{\hide}[1]{}

\newcommand{\etal}{{\textit{et~al.}}}
\newcommand{\ie}{i.e.,}
\newcommand{\eg}{e.g.,}

\tcbset{top=0mm,bottom=0mm,left=0mm,right=0mm,fonttitle=\bfseries\scriptsize\color{gray},colbacktitle=white,enhanced,attach boxed title to top center={yshift=-1mm},boxed title style={top=0mm,bottom=0mm,left=0mm,right=0mm}}

\begin{document}

\title[\methodName: An Optimization-based Post-processing for Graphic Design Layout Generation]%
      {\methodName: An Optimization-based Post-processing for Graphic Design Layout Generation}

\author[paper1002]{paper1002}

\author[I-Chao Shen et al.]
{\parbox{\textwidth}{\centering I-Chao Shen$^{1}$\orcid{0000-0003-4201-3793} ~ Ariel Shamir$^{2}$\orcid{0000-0001-7082-7845} ~ Takeo Igarashi$^{3}$\orcid{0000-0002-5495-6441}
        }
        \\
{\parbox{\textwidth}{\centering $^1$ ichaoshen@g.ecc.u-tokyo.ac.jp , The University of Tokyo, Japan \\
$^2$ arik@runi.ac.il, Reichman University, Israel \\
$^3$ takeo@acm.org, The University of Tokyo, Japan
      }
}
}

\teaser{
  \centering
  \includegraphics[width=\textwidth]{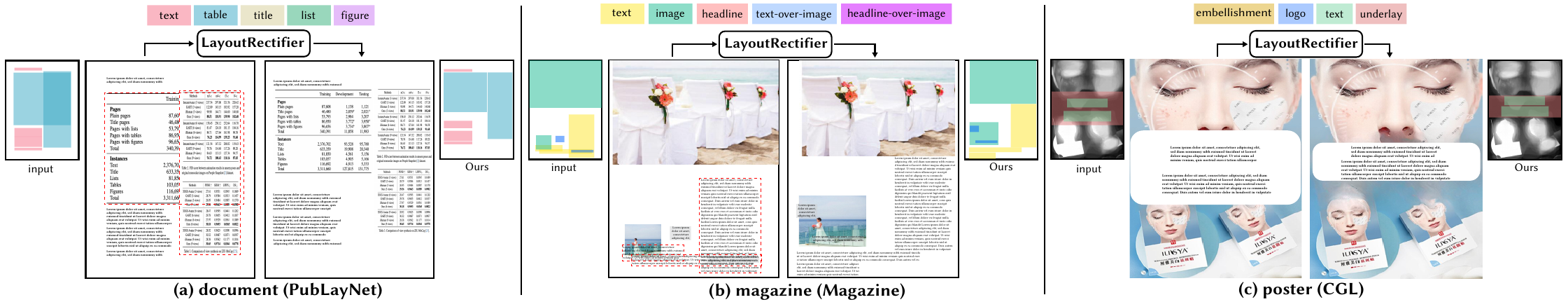}
  \caption{
    \methodName rectifies the layouts generated by various learning-based layout generation methods.
    For (a) document layouts, it improves the alignment and prevents unwanted overlap between layout elements. 
    For (b) magazine layouts, our method further improves the desired containment.
    For (c) poster layouts, our method also reduces occlusion on salient image content around the middle part.
    Note that the content in these layouts are for visualization purpose only.
    (\protect \redline: unwanted overlaps.)
  }
  \label{fig:teaser}
}



\maketitle

\begin{abstract}
Recent deep learning methods can generate diverse graphic design layouts efficiently.
However, these methods often create layouts with flaws, such as misalignment, unwanted overlaps, and unsatisfied containment.
To tackle this issue, we propose an optimization-based method called \methodName, which gracefully rectifies auto-generated graphic design layouts to reduce these flaws while minimizing deviation from the generated layout.
The core of our method is a two-stage optimization.
First, we utilize grid systems, which professional designers commonly use to organize elements, to mitigate misalignments through discrete search.
Second, we introduce a novel box containment function designed to adjust the positions and sizes of the layout elements, preventing unwanted overlapping and promoting desired containment. 
We evaluate our method on content-agnostic and content-aware layout generation tasks and achieve better-quality layouts that are more suitable for downstream graphic design tasks.
Our method complements learning-based layout generation methods and does not require additional training.
\end{abstract}

\section{Introduction}
\label{sec:intro}
Creating a layout is a central problem for various graphic design tasks, such as documents, magazines, and posters.
However, creating layouts is a complex and time-consuming task that requires design expertise.
To reduce the efforts involved in creating layouts, researchers continuously explore approaches to automatically generate layouts~\cite{inoue2023layoutdm,kikuchi2021constrained,kong2022blt,Lin_2023_ICCV,li2018layoutgan,o2014learning,zhang2023layoutdiffusion,cheng2023play, zhengsig19,horita2024retrievalaugmented} 
under \sighl{some layout-specific criteria.}
\sighl{For example, overlapping elements are disallowed in a document layout, but they are common in a magazine layout.}

\begin{figure}[t]
  \centering
  \includegraphics[width=\linewidth]{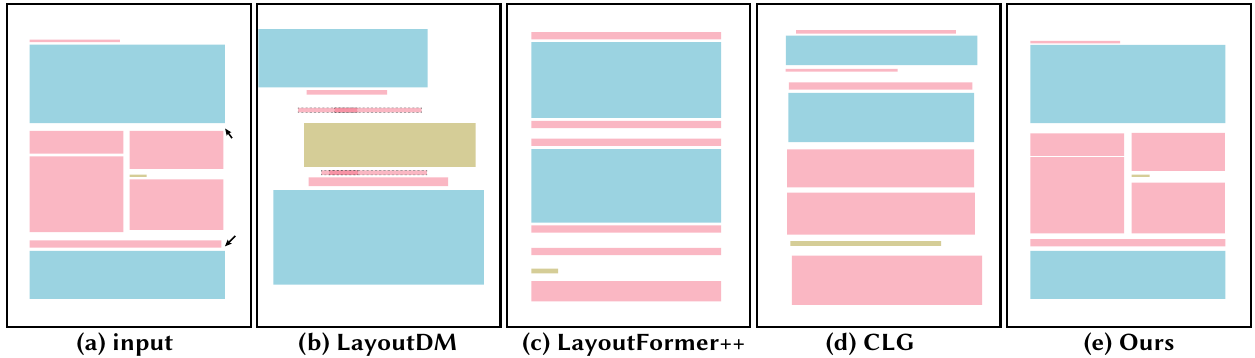}
  \caption{
   (a) The input layout generated by a learning-based layout generation method only has minor misalignment issue (the black arrow region) without undesired overlap. Existing layout refinement methods~\cite{inoue2023layoutdm,Jiang_2023_CVPR, kikuchi2021constrained}  either (b,c,d) introduce additional flaws or deviate severely from the original layout. (e) Our method gracefully correct the misalignment and successfully preserve the layout structure.
  }
  \label{fig:preserve_fig}
\end{figure}

Many methods use a learning-based approach for creating latyouts. \sighl{The main advantage of these learning-based methods is that they can generate diverse layouts efficiently.
However, these generated layouts often include common flaws that do not comply with the layout-specific criteria.} 
First, the generated graphic elements in the layout are often misaligned with each other.
Second, there are often undesired overlapping relations between layout elements.
Third,  unsatisfied containment between elements often happens in layouts with many images and texts, such as magazine and poster.
These flaws occur because existing layout generation methods learn to create layouts solely by modeling the attribute distributions of the training datasets.
As a result, it becomes challenging to provide explicit guidance to avoid these flaws.
While there are some existing learning-based layout refinement methods, they often introduce new flaws or destroy the original layout structure (\mbox{\autoref{fig:preserve_fig}}).
\pghl{Additionally, existing optimization-based layout generation methods are either tailored for different domains, such as interior design~\mbox{\cite{merrell2011interactive,yu2011make}}, or requires a good example to learn the relationships between elements~\mbox{\cite{o2014learning}}.}
Therefore, it is essential to develop a method that addresses common flaws found in layouts generated by recent learning-based methods while preserving the diversity of layouts generated by these methods.

In this paper, we propose \methodName: an optimization-based method to fulfill this need.
Our method initially identifies alignment relations of the input layout using perceptual criteria~\cite{koffka2013principles,hess1999integration}.
Meanwhile, our method utilizes an exemplar grid system~\cite{müller1981grid} to enhance element alignment.
\sighl{
Given a flawed layout and the exemplar grid, our method aims to adjust element parameters to enhance alignment by adhering to the exemplar grid and mitigate overlapping issues.
This goal requires solving an discrete-continuous problem: we need to search a grid cell for an element to snap to and adjust its continuous parameters.
Therefore, we solve this problem in two alternating stages. 
}

\sighl{First, we adopt a discrete search-and-snap strategy to encourage each layout element to align with the close-by grid lines of the reference grid, minimizing misalignments and other criteria, such as overlapping and occlusion.}
\sighl{Second, our method continuously updates the parameters of elements to address overlapping and containment issues using a novel differentiable box containment function.}
Unlike existing functions such as Intersection-over-Union (IoU) and its variations~\cite{rezatofighi2019generalized,zheng2020distance,zheng2021enhancing}, our box containment function can define gradients for optimization even when elements are completely contained or totally separate from each other.
\sighl{
The discrete search stage adjusts the layout locally, while the continuous updates allow the elements to move further to resolve unwanted overlaps and containment issues.
}

We evaluate our method on both content-agnostic and content-aware layout generation tasks using document, magazine, and poster layouts.
The results show that our method can resolve layout problems effectively across various types of layouts generated by various learning-based methods.
Importantly, our method complements existing and future learning-based layout generation methods and doesn't require any additional training process.

\section{Related Work}
\label{sec:related}
\subsection{Graphic Layout Generation and Refinement}
Automatic layout generation can help simplify the process of creating graphic design, especially for novices.
Initially, research focused on generating and refining a layout from itself~\cite{designspace,o2014learning}.
Later, data-driven methods gained more attention for generating diverse layouts.
The development of these methods was based on various datasets across graphic design domains, including scientific paper~\cite{zhong2019publaynet}, magazine~\cite{zhengsig19}, and poster~\cite{Hsu_2023_CVPR,cgl2022}.
These data-driven layout generation methods use various deep learning models, such as VAE~\cite{lee2020neural}, GAN~\cite{kikuchi2021constrained,li2018layoutgan}, autoregressive models~\cite{kong2022blt,Lin_2023_ICCV}, and diffusion models~\cite{inoue2023layoutdm,cheng2023play,zhang2023layoutdiffusion,hui2023unifying}, to tackle both unconditional and conditional layout generations.
Most recently, large language models were also found to be useful for layout generation abilities~\cite{lin2024layoutprompter}.
While these methods can also perform layout refinement, the refined layouts using them often deviate too much from the original layout and fail to resolve problems in it as shown in~\autoref{fig:preserve_fig}.


\subsection{Grid System and Snapping}
The grid system is a spatial organization principle~\cite{müller1981grid} has been widely adopted in designing various printed and digital media layouts, such as webpages, advertisements, and presentations.
Dayama~\etal~\shortcite{dayama2020grids} proposes an optimization method for generating and exploring grid-based layouts.
However, their method has limitations in dealing with overlapping elements and restricted to the layouts with rectangular outlines.
In contrast, our method can generate refined layouts with desired overlaps and element arrangement according to reference white spaces.

\subsection{Bounding Box Metric and Cost Function}
Intersection over Union (IoU) is a widely used metric for assessing the accuracy of bounding box in object detection and layout generation.
There are several variants of IoU that incorporates geometric factors, \eg~bounding area~\cite{rezatofighi2019generalized}, central point distance~\cite{zheng2020distance}, and aspect ratio~\cite{zheng2021enhancing}.
These variants address the gradient deficiency issue in cases where areas do not overlap, thus accelerating convergence speed.
However, these cost functions are designed to increase the overlap between boxes, which means they cannot be applied to avoid overlap in a unified formulation.
Furthermore, incorporating additional geometric factors does not always yield the desired overlap in layout generation task.  
Instead, our containment cost function can increase and decrease overlaps between boxes in a unified formulation without the gradient deficiency issue.
\sighl{Minar\v{c}\'{\i}k~\etal~\mbox{\cite{Minarcik2024Minkowski}} proposed Minkowski penalty for handling general 2D shapes arrangement.
On the contrary, we focus on refining graphic design layout using the grid system. 
}
\begin{figure*}[h]
  \centering
  \includegraphics[width=\linewidth]{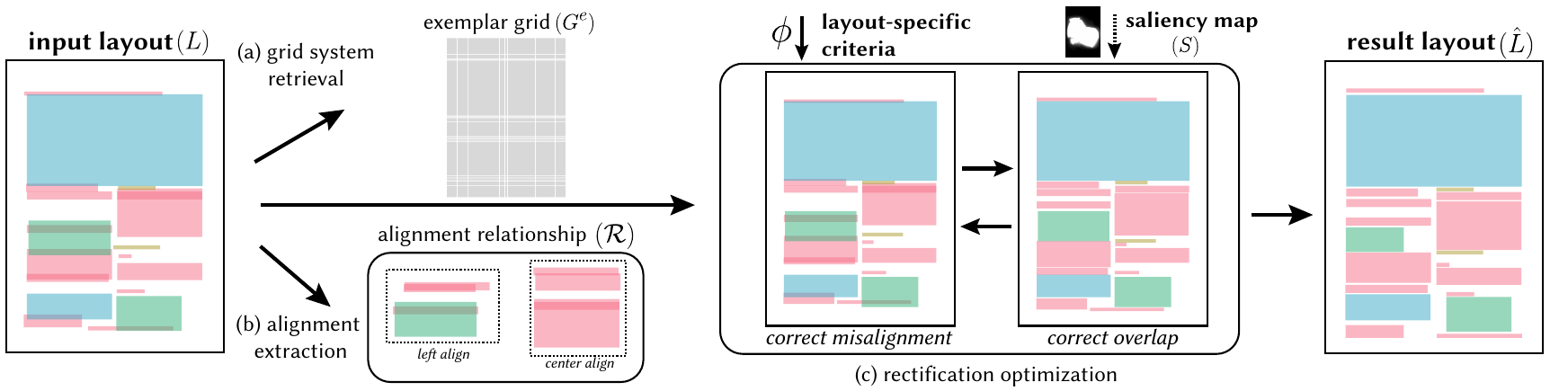}
  \caption{
\textbf{Overview of our method.}
Given an input layout and layout-specific criteria, our method first (a) retrieves an exemplar grid and (b) extracts alignment relations. 
Then, (c) we obtain the result layout by iteratively enhance alignment and mitigate overlapping issues through an iterative rectification optimization.
}
  \label{fig:overview}
\end{figure*}
\section{Method}
\subsection{Overview and Layout Representation}
As shown in~\autoref{fig:overview}, our refinement method takes two inputs: a layout $L$ and a layout-specific criteria $\boldsymbol{\phi}$ specific to the layout.
\sighl{
To refine a layout from a content-aware generation task, we use an additional input -- the saliency map $S$ of the canvas image $I$. This map is obtained by an off-the-shelf saliency detection method~\mbox{\cite{qin2022highly}}. 
}

The input layout $L$ is generated by content-agnostic~\cite{kikuchi2021constrained,inoue2023layoutdm, kong2022blt, Jiang_2023_CVPR} or content-aware~\cite{horita2024retrievalaugmented} learning-based layout generation methods.
We denote the layout by $L= \{(c_1, \mathbf{b}_1), ..., (c_N, \mathbf{b}_N)\}$, where $c_i \in \{1,...,C\}$ indicates the category of the $i$-th element, $b\in [0,1]^4$ indicates the bounding box in normalized coordinates, and $N$ is the number of element in the layout.
The $i$-th element box $b_i$ is represented as $[x_i, y_i, w_i, h_i]$, where $(x_i,y_i)$ represents its center location and $(w_i, h_i)$ represents its width and height.
Given the input layout $L$, we will retrieve an exemplar grid $G^{e}$ extracted from an exemplar layout from the reference layout dataset (\autoref{fig:overview}(a)).
Additionally, our method will extracts a set of intrinsic alignment relations $\mathcal{R}$ between elements (\autoref{fig:overview}(b)).

\sighl{The layout-specific criteria includes three sets of element types:} 
$\boldsymbol{\phi}=(\phi_{\text{child}}, \phi_{\text{parent}}, \phi_{\text{others}})$,
\sighl{which are the \textit{child} set $\phi_{\text{child}}$, the \textit{parent} set $\phi_{\text{parent}}$, and the \textit{others} set $\phi_{\text{others}}$.
Containment is only allowed when an element belonging to $\phi_{\text{child}}$ is overlayed on an element belonging to $\phi_{\text{parent}}$.
For example, when dealing with a document layouts element overlay is totally prohibited, so all element types are in $\phi_{\text{others}}$.
On the other hand, when dealing with magazine layouts, the types of element ``text-over-image'' and ``headline-over-image'' can overlay on the ``image'' element.
In this case, ``image'' belongs to $\phi_{\text{parent}}$, ``text-over-image'' and ``headline-over-image'' belong to $\phi_{\text{child}}$, and ``text'' and ``headline'' belong to $\phi_{\text{others}}$.
}

Taking these into account, we perform a two-stage rectifying optimization  (\autoref{fig:overview}(c)) that enforces the alignment of the layout elements with the reference grid $G^{e}$ based on the alignment relationships $\mathcal{R}$.
\sighl{In the first stage, we perform an discrete search using the reference grid $G^{e}$ to mitigate misalignment and also resolve the undesired overlay and improve the desired containment using a box containment function to obtain the final layout $\hat{L}$.}

In our work, we retrieve $M=5$ exemplar layouts from the corresponding training layout dataset using Intersection over Union (IoU) similarity and utilize the grid systems extracted from them as reference grids.
Then, we generate $M$ rectified layouts and select the one with the fewest flaws as the final result.
The purpose of creating multiple rectified layouts is to ensure that even if one reference grid does not match the input layout, our method can still mitigate the flaws.
See~\autoref{alg:repair} for the detailed process.
\begin{algorithm}
\caption{
\methodName refinement process}
\label{alg:repair}
\begin{algorithmic}[1]
\INPUT input layout $L$, a layout dataset $\mathbf{L}$, number of candidates $M$.
\OUTPUT layout with the least flaws $\hat{L}$.

\State $\mathbf{G}^{e} \gets \text{retrieve}(L, \mathbf{L}, M)$ 
\State $\mathcal{R} \gets \text{detect-alignment}(L)$
\State $\hat{L} \gets \emptyset$
\For{$G^{e}_{k} \in \mathbf{G}^{e}$}
    \Statex \LeftComment{1}{$T$: max iteration number}
    \For{$t \gets 1$ to $T$} 
        \State $L_t \gets \text{search-and-snap}(L_t, G^{e}_{k}, \mathcal{R})$ 
        \Statex \LeftComment{2}{$L_t$: refined layouts after at $t$-th iteration}
        \State $L_t \gets \text{minimize } (\mathcal{E}_{\text{ove}}+\mathcal{E}_{\text{contain}}+\lambda_{\text{aspect}}\mathcal{E}_{\text{aspect}}+\lambda_{\text{area}}\mathcal{E}_{\text{area}}$) 
    \EndFor
\If{$\text{score}(L_{T}) < \text{score}(\hat{L}$)}
    \State $\hat{L} \gets L_{T}$
\EndIf
\EndFor
\State \textbf{return} $\mathbf{\hat{L}}$
\end{algorithmic}
\end{algorithm}

\subsection{Grid System}
\label{sec:gridsys}
The grid system is a framework that designers use to organize elements and ensure alignment between them~\cite{müller1981grid}.
It helps organizing elements and results in more balanced, organized, and visually appealing layouts.
In this work, we assume that most high-quality graphic designs are created with the help of various grid systems, and we can leverage these grid systems to refine a layout containing flaws.
However, in a public layout dataset, we typically only have access to the final layouts, not the grid systems used during the design process. 
Therefore, we instead estimate a grid system that best describes each layout.

\sighl{A grid system consists of grids of varying sizes surrounded by margins and separated by gutters, which are spaces between the grids that prevent a layout from becoming overcrowded} (\autoref{fig:gridsys}(d)).
\sighl{This type of grid system organizes elements in a structured manner based on their level of importance, making it easier for users to navigate and interpret complex information or designs through a clear visual or logical structure.}

\subsubsection{Constructing a grid system}
\begin{figure}[t]
  \centering
  \includegraphics[width=\linewidth]{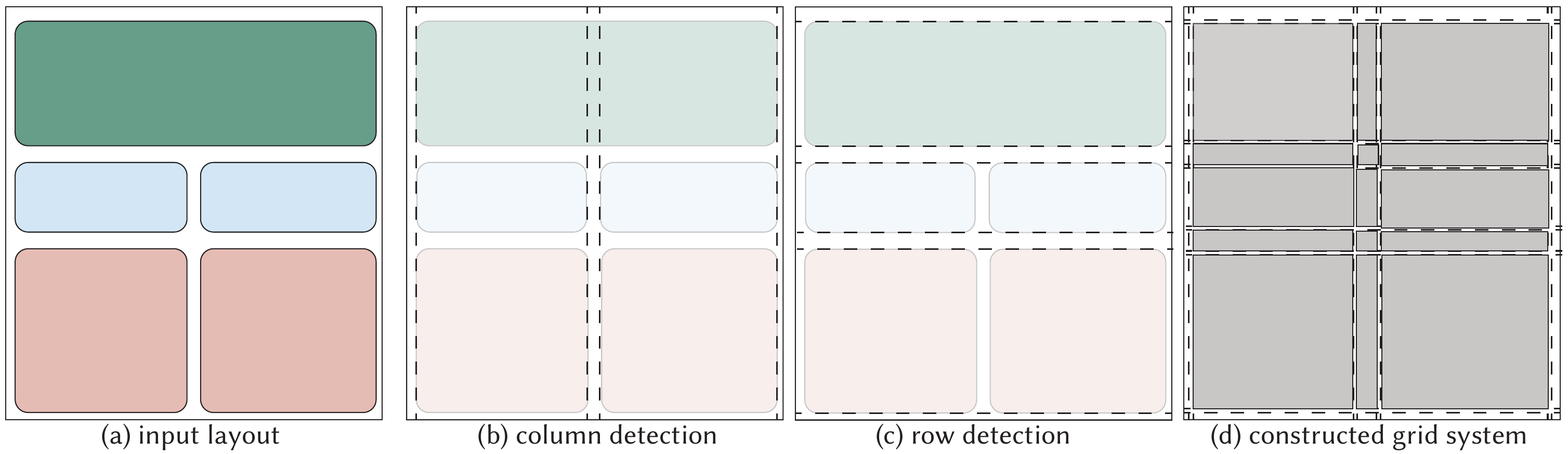}
  \caption{
  Given (a) an input layout, we first detect initial (b) column and (c) row grid locations using each corners of element boxes. 
  (d) We divide each initial column/row grid locations and obtain the final grid system.
  }
  \label{fig:gridsys}
\end{figure}

Our goal is to construct a grid system $G$ that accurately represents a given layout $L$ from a layout dataset $\mathbf{L}$.
We start by identifying the smallest left and top corners of all element boxes in the layout dataset $\mathbf{L}$ to determine the left and top margins of the grid system.
Similarly, we identify the largest bottom and right corners of all element boxes in $\mathbf{L}$ as the right and bottom margins.
\rvhl{These corners are used to define a global boundary across all grids.
This is necessary due to the varying spatial distribution of layout elements across different layouts. To accommodate this variation, we use this global boundary as a common reference to construct grids that can represent diverse layouts.
}
Next, \rvhl{within this global boundary,} we use the left and right values of all element boxes in $L$ as initial column locations (\autoref{fig:gridsys}(b)) and the top and bottom values of all element boxes in $L$ as initial row locations (\autoref{fig:gridsys}(c)) to construct $G$.
\sighl{To create the gutters between columns,} we create a separate grid with a width equal to the original width minus two gutter values. 
We perform the same operation on the rows and obtain the final grid system (\autoref{fig:gridsys}(d)).
We provide examples of constructed grid systems as supplemental material.

\subsection{Alignment Extraction}
\label{sec:align}
Alignment in graphic design refers to the arrangement of elements in a layout, either in relation to one another or a common baseline.
The goal is to create order and improve element grouping. 
Most existing data-driven methods only implicitly encourage alignment by generating layouts that resemble those in the training data.
LayoutGAN~\cite{li2020attribute} proposes an alignment loss function to explicitly improve alignment during training.
Unfortunately, this loss function can produce undesirable results as it encourages alignment between elements that are not perceptually associated.
As a result, the generated layout may not have the desired perceptual hierarchy, even though this loss function is minimized.

\begin{figure}[t!]
  \centering
  \includegraphics[width=\linewidth]{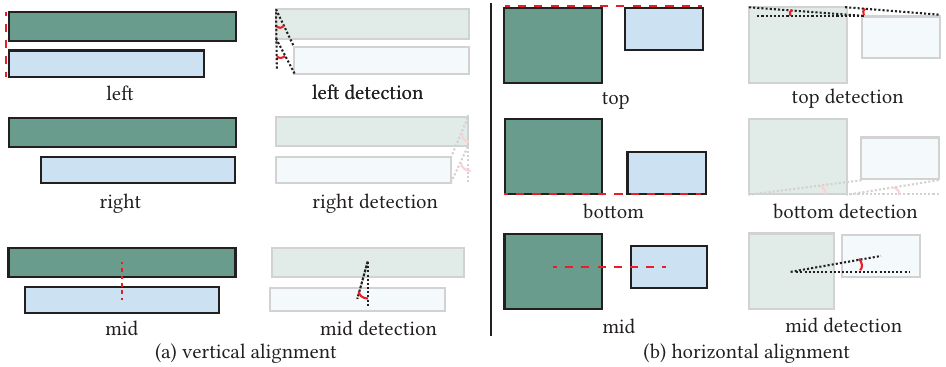}
  \caption{
  A subset of (a) vertical and (b) horizontal alignment categories used in our method and how they are detected. 
  }
  \label{fig:small_alignment}
\end{figure}

To address this issue, we propose using perceptual criteria to identify the intrinsic alignment relations between different element boxes in the generated layout.
Given two element boxes $b_i$ and $b_j$, our method determines whether they fit into any of the alignment categories commonly used in graphic design, namely:
\begin{itemize}
    \item \textbf{vertical alignment}: left, right, vertical mid, left-right (\autoref{fig:small_alignment}(a)).
    \item \textbf{horizontal alignment}: top, bottom, horizontal mid, top-bottom (\autoref{fig:small_alignment}(b)).
\end{itemize}

We utilize Gestalt continuity criteria to determine whether two boxes fit into any of the alignment categories listed above.
\rvhl{We build upon previous work~\mbox{\cite{bessmeltsev2015modeling}} and define two boxes as being horizontally (or vertically) aligned if the angle between the line connecting their endpoints and a horizontal (or vertical) line is less than $18\degree$. 
This threshold is based on the finding from~\mbox{\cite{hess1999integration}} that show over $90\%$ of viewers perceive disconnected curve segments as a single contour when the angle is below this limit.
}

\subsection{Box Containment Function}
Previous learning-based methods avoid element overlay using reconstruction loss~\cite{kikuchi2021constrained} or overlap loss~\cite{li2020attribute}.
However, these methods face issues such as lack of explicit guidance or \sighl{the gradient deficiency issues: no gradient is defined when an element is entirely contained by another element or when they do not overlap}.
To address these issues, we propose a box containment cost function that prevents overlay or enforce containment \sighl{without gradient deficiency issues.}

Our box containment function is based on the distance IoU metric (DIoU)~\cite{zheng2020distance}, which is defined for two elements $b_c, b_p$ in the generated layout as: 
\begin{align}
E_{\text{DIoU}} = 1 - \tcboxmath[colback=white,colframe=purpleD,title=IoU]{\frac{|b_c \cap b_p|}{|b_c \cup b_p|}} + \tcboxmath[colback=white,colframe=greenD,title=distance penalty]{\frac{\rho(b_c, b_p)^2}{c^2}},
\end{align}
where the term $\rho(b_c, b_p)$ denotes the distance between the centers of $b_c$ and $b_p$, and $c$ represents the diagonal distances of the bounding box of $b_c$ and $b_p$.
\sighl{
This term ensures that gradients exist even when there is no overlap.
}

Despite $E_{\text{DIoU}}$ addressing the gradient deficiency issues, it still has two problems that hinder its effectiveness as a cost function for improving undesired overlay and containment.
First, our goal is to ensure that a box is entirely contained by or has no overlay with another box.
Thus, it is ineffective to use the area of the union of both boxes.
Second, $E_{\text{DIoU}}$ aims to align the centroids of two boxes.
However, the alignment of centroids will produce undesired layouts, \eg~when we want left-aligned or right-aligned boxes.

We propose two solutions to address these two issues.
\sighl{First, we replace the IoU term used in DIoU with an ``Intersection over Child Area'' term (IoCA), which is defined as $\frac{|b_c \cap b_p|}{|b_c|}$, where $b_p$ represents an element in the parent set $\phi_{\text{parent}}$ and $b_c$ represents an element in the child set $\phi_{\text{child}}$.}
Second, we introduce a weight value $w_d^{+}=1.0-\text{IoCA}$ to control the effect of the distance term. 
$w_d^{+}$ decreases with an increase in IoCA value.
Combining these, our \textit{positive} box containment cost function is defined as:
\begin{align}
E^{+}_{\text{contain}}= 1 - (\text{IoCA} - w_d^{+}\frac{\rho(b_c, b_p)^2}{c^2}).
\label{eq:pos_contain}
\end{align}
Note that $E^{+}_{\text{contain}}$ becomes zero once $b_c$ is entirely contained by $b_p$.

\begin{figure}[t]
  \centering
  \includegraphics[width=\linewidth]{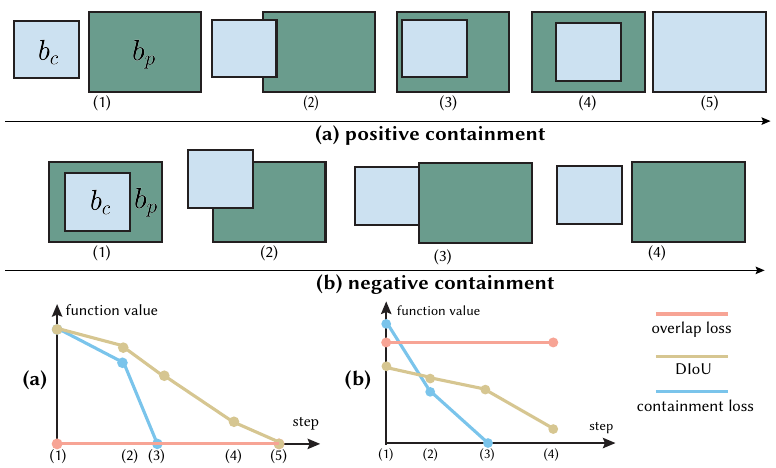}
  \caption{
  Illustration of the optimization process using (a) positive and (b) negative box containment functions.}
  \label{fig:cont_loss_illustration}
\end{figure}

This optimization process is illustrated in ~\autoref{fig:cont_loss_illustration}(a) (blue line).
On the contrary, with the existing overlap loss~\cite{li2020attribute} (red line), $b_c$ remains unchanged because of lack of gradients.
On the other hand, with $\text{DIoU}$ (brown line), the centroids and sizes of both boxes will be forced to align.

To avoid overlap between $b_c$ and $b_p$, we define the \textit{negative} box containment cost function as:
\begin{align}
E^{-}_{\text{contain}} = 1 - (\text{IoCA} + w_d^{-}\frac{\rho(b_c, b_p)^2}{c^2}),
\label{eq:neg_contain}
\end{align}
where $w_d^{-} = \text{IoCA}$. 
The maximum value of this function occurs when two centroids aligned and will gradually decrease as $b_c$ moves away from $b_p$.
The distance term will diminish when $b_c$ has no overlap with $b_p$.
This optimization process is illustrated in~\autoref{fig:cont_loss_illustration}(b) (blue line).
On the contrary, with the existing overlap cost function (red line), $b_c$ will remain contained by $b_p$.
If we try to minimizing $-E_{\text{DIoU}}$ (brown line), $b_c$ will be forced to move away from $b_p$ to minimize the distance term.
\sighl{It is important to note that, it is nontrivial to use our differentiable cost function into recent learning-based methods since they require a non-differentiable decoding step to obtain the element parameters.
}

\subsection{Iterative Rectification Optimization.}

To refine the input flawed layout $\mathcal{E}$ using the detected alignment relations $\mathcal{R}$ and an reference grid $G^{e}$, we introduce our objective function and the optimization technique.
\subsubsection{Energy Function.}
Our total energy function $\mathcal{L}_{\text{all}}$ consists of \sighl{six} components: 
\begin{align}
\mathcal{E}_{\text{all}}=\mathcal{E}_{\text{align}} &+\mathcal{E}_{\text{dist}}+ \mathcal{E}_{\text{ove}} + \mathcal{E}_{\text{cont}}+ \lambda_{\text{aspect}}\mathcal{E}_{\text{aspect}} + \lambda_{\text{size}}\mathcal{E}_{\text{size}},
\label{eq:opt_obj} 
\end{align}
where we set $\lambda_{\text{aspect}}=10$ and $\lambda_{\text{size}}=100$ throughout all experiments.
Following we introduce the definition of each energy term.

\begin{description}[leftmargin=0pt]
\item[Alignment term ($\mathcal{E}_{\text{align}}$)]
This alignment term aims to enforce ``paired'' and ``unpaired'' alignments.
In the case of ``paired'' alignment, if an element box $b_i$ is included in $\mathcal{R}$, meaning it aligns with at least one element box, we use the following term to enforce this alignment relation to occur:
\begin{align}
E_{\text{align}}^{\text{paired}}(b_i) &= 
\begin{cases}
\sum_{j \in \mathcal{R}(i)} |x^{*}_i-x^{*}_j|, & \text{if } (i,j) \text{ is vertical alignment.}\\
\sum_{j \in \mathcal{R}(i)} |y^{*}_i-y^{*}_j|, & \text{if } (i,j) \text{ is horizontal alignment.}\\
\end{cases}
\end{align}
Here, ``*=(L, R, C)'' denotes the left, right, and center coordinates for vertical alignment, while ``*=(T, B, C)'' denotes the top, bottom, and center coordinates for horizontal alignment.
Otherwise, we use the unpaired alignment loss introduced in~\cite{li2020attribute} as our $E_{\text{align}}^{\text{unpaired}}$.
It measures a similar alignment loss for both vertical and horizontal directions between each element pairs even without perceptual association, 


\item[Overlap term ($\mathcal{E}_{\text{ove}}$)]
The overlap term is defined by calculating the \textit{negative} containment cost function between element boxes, except when one of the box categories belongs to $\phi_{\text{parent}}$:
\begin{align}
\mathcal{E}_{\text{ove}}(L) &= \sum_{i=0}^{N} \sum_{j \neq i} E^{-}_{\text{contain}}(b_i, b_j)\sigma^{-}(b_i, b_j), \\ \nonumber
\sigma^{-}(b_i, b_j) &= 
\begin{cases}
0.0, & \text{if } (c_i \in \phi_{\text{parent}} \text{ and } c_j  \notin \phi_{\text{parent}}),\\
0.0, & \text{if } (c_i \notin \phi_{\text{parent}} \text{ and } c_j  \in \phi_{\text{parent}}),\\
1.0, & \text{otherwise.} \\
\end{cases}
\end{align}

\item[Containment term ($\mathcal{E}_{\text{cont}}$)]
We minimize the \textit{positive} containment function between elements only when one box category belongs to $\phi_{\text{parent}}$:
\begin{align}
\mathcal{E}_{\text{contain}}(L) &= \sum_{i=0}^{N} \sum_{j\neq i} E^{+}_{\text{contain}}(b_i, b_j)\sigma^{+}(b_i, b_j), \label{eq:containment_term} \\ \nonumber
\sigma^{+}(b_i, b_j) &= 
\begin{cases}
1.0, & \text{if } (c_i \in \phi_{\text{parent}} \text{ and } c_j  \notin \phi_{\text{parent}}),\\
1.0, & \text{if } (c_i \notin \phi_{\text{parent}} \text{ and } c_j  \in \phi_{\text{parent}}),\\
0.0, & \text{otherwise.} \\
\end{cases}
\end{align}

\item[Aspect ratio preservation term ($\mathcal{E}_{\text{aspect}}$)]
\sighl{For each layout element, this term is defined by calculating the deviation of its aspect ratio in the refined layout and the input layout:}
\begin{align}
\mathcal{E}_{\text{aspect}} = \sum_{i\in L_{\text{aspect}}}(\frac{w_i}{h_i}-\frac{\hat{w}_i}{\hat{h}_i})^2,
\end{align}
\sighl{
where $L_{\text{aspect}}$ is the set of layout element whose aspect ratio should be preserved, and ($\hat{w}_i$, $\hat{h}_i$) are the original width and height of element in the input layout. 
}

\item[Size preservation term ($\mathcal{E}_{\text{size}}$)]
\sighl{For each layout element, this term is defined by calculating the width and height differences before and after refinement:}
\begin{align}
\mathcal{E}_{\text{size}} = \sum_{i\in L_{\text{area}}}({w_i}-\hat{w}_i)^2+({h_i}-\hat{h}_i)^2,
\end{align}
\sighl{
where $L_{\text{size}}$ is the set of element whose size should be preserved.
}

\end{description}

\subsubsection{Solving the Optimization Problem}
\label{sec:opt_strategy}
\sighl{
Optimize~\mbox{\autoref{eq:opt_obj}} while strictly adhering to the exemplar grid $G^{e}$ is nontrivial.
This is because, for each element, we need to discretely determine which cell in $G^{e}$ to snap to in order to enhance the alignment and adjust its parameters continuously to avoid overlapping. 
}
To address this challenge, we obtain a solution by alternating between two stages:

    \begin{figure}
\centering
\subfloat[Snap option examples.]{{\includegraphics[width=.98\linewidth]{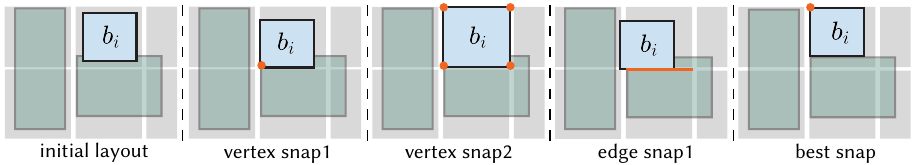} }\label{fig:sas_example}}

\subfloat[Two-stage strategy example.]{{\includegraphics[width=.98\linewidth]{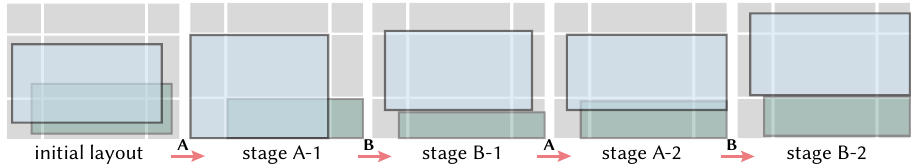} }\label{fig:two_stage_toy}}

\caption{(a) Element box $b_i$ can snap by vertex or edge. Our method selects the best snap option that avoids both overlap and misalignment.
  (orange point and line indicate snap vertex and edge.)
  (b) Given the initial layout, our method enhances misalignments and mitigates overlaps by alternating stage A and B.
  }
\label{fit:two_stage_illustration}
    \end{figure}

\begin{description}[leftmargin=0pt]
\item[Stage A: rectify misalignment via discrete search.]
To correct misalignment while adhering to the reference grid $G^{e}$, we perform a \textit{search-and-snap} process aimed at minimizing $\mathcal{E}_{\text{all}}$ (\autoref{eq:opt_obj}).
\sighl{We illustrate the search-and-snap process in~\mbox{\autoref{fig:sas_example}}.
Specifically, for each box $b_i$, we examine various commonly used snapping options to search the one that results in the minimum $\mathcal{E}_{\text{all}}$ (\mbox{\autoref{eq:opt_obj}}).
The key idea is that $\mathcal{E}_{\text{all}}$ encourages $b_i$ to snap to the nearest grid while adhering to alignment and other requirements regarding other boxes.
When rectifying layouts generated from the content-aware task, we also minimize an additional occlusion term $\mathcal{E}_{\text{occ}}$ to ensure that the snapped box occludes the content as little as possible.
See supplement for the details of the box distance term $\mathcal{E}_{\text{dist}}$, the occlusion term $\mathcal{E}_{\text{occ}}$, and the search-and-snap process.
}

\item[Stage B: rectify overlap via continuous update.]
For the second stage, to rectify any overlay and containment issues that may arise \sighl{while preserving elements' aspect ratio or size}, we minimize the following function using Adam~\cite{adam} for $100$ iteration:
\begin{align}
\mathcal{E}_{\text{ove}}+\mathcal{E}_{\text{contain}}+\lambda_{\text{aspect}}\mathcal{E}_{\text{aspect}}+\lambda_{\text{size}}\mathcal{E}_{\text{size}}.
\end{align}

\begin{figure*}[h!]
  \centering
  \includegraphics[width=\linewidth]{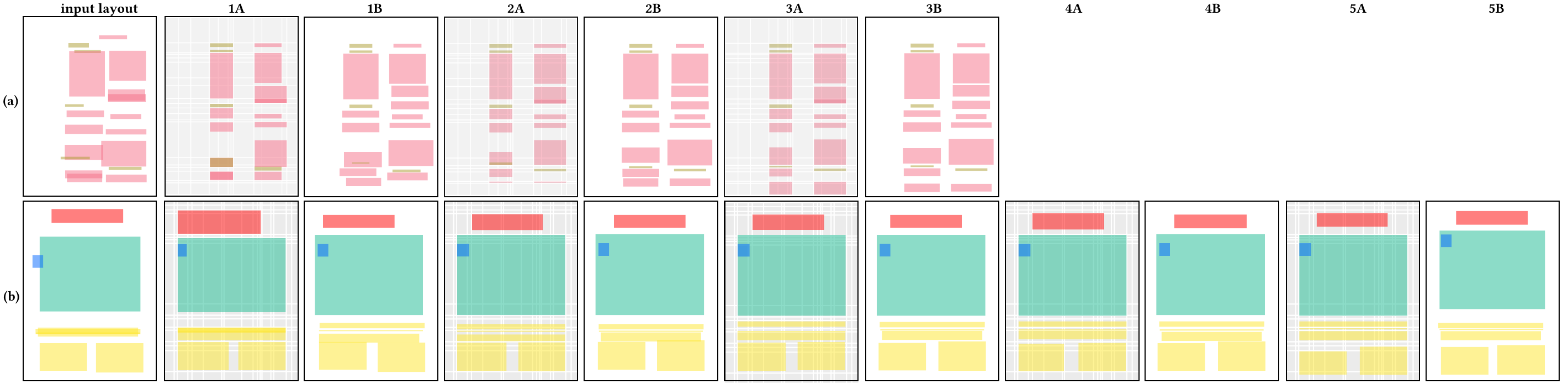}
  \caption{
  \textbf{Optimization process illustration.}
  (a) Example optimization processes for (a) a document layout and (b) a magazine layout.
  For the document layout, our method rectify all flaws after three iterations.
  }
  \label{fig:opt_process}
\end{figure*}

\begin{table*}
\small
\centering
\begin{tabular}{c|ccccc|ccccc|cccc}
\toprule
\multicolumn{1}{c}{} & \multicolumn{5}{c}{\textbf{Ove$\downarrow$}} & \multicolumn{5}{c}{\textbf{Align (x100)$\downarrow$}} & \multicolumn{4}{c}{\sighl{\textbf{Sim$\uparrow$}}} \\
\cmidrule(lr){2-6} \cmidrule(lr){7-11} \cmidrule(lr){12-15}
 & {Ori} & \sighl{LDM*} & \sighl{LF++*} & \sighl{LP} & {Ours} & {Ori} & \sighl{LDM*} & \sighl{LF++*} & \sighl{LP} & {Ours} & \sighl{LDM*} & \sighl{LF++*} & \sighl{LP} & {Ours}\\
\midrule
LGAN++  & 0.108 & 0.175 & 0.019 & \seccell{0.018} &\bestcell{0.0014} & 0.18 & 0.156 & \seccell{0.036} & 0.176 & \bestcell{0.027} & 0.716 & \seccell{1.551} & 0.648 & \bestcell{1.942} \\
BLT  & 1.049  & \seccell{0.185} & 0.446 & 0.40 & \bestcell{0.153} & 0.139 & 0.175 & \bestcell{0.022} & 0.645 &\seccell{0.053} & 0.751 & \seccell{1.702} & 0.417 &\bestcell{1.951} \\
LDM & 0.155 & 0.168 & \seccell{0.021} & 0.146 & \bestcell{0.003} & 0.12  & 0.166 & \bestcell{0.019} & 0.021 & \seccell{0.020}  & 0.794 & \seccell{1.606} & 0.357 &  \bestcell{2.21}\\
LF++ & 0.009 & 0.20 & \seccell{0.004} & 0.13 & \bestcell{0.002} & 0.023 & 0.44 & 0.013 & \seccell{0.007} & \bestcell{0.002} & 0.30 & 0.389 & \seccell{0.695} & \bestcell{2.113} \\
\bottomrule
\end{tabular}
\caption{
\textbf{Quantitative comparison on content-agnostic generation (PubLayNet dataset).}
``Ori'' denotes the layouts generated by each learning-based generation method.
\rvhl{``LDM*'' and ``LF*'' denote the refinement models based on their original generation model, respectively.}
We highlight the \besthint{first} and \sechint{second} best result for each metric and each layout generation method.
Our method successfully achieve better improvements on overlay and alignment while preserving the similarity between the input layouts and the rectified layouts.
}
\label{tab:publaynet_quan}
\end{table*}

\begin{table*}
\small
\centering
\begin{adjustbox}{width=\linewidth, center}
\begin{tabular}{c|ccccc|ccccc|ccccc|cccc}
\toprule
\multicolumn{1}{c}{} & \multicolumn{5}{c}{\textbf{Ove$\downarrow$}} & \multicolumn{5}{c}{\textbf{Align (x100)$\downarrow$}} & \multicolumn{5}{c}{\textbf{Cont$\uparrow$}} & \multicolumn{4}{c}{\sighl{\textbf{Sim$\uparrow$}}}\\
\cmidrule(lr){2-6} \cmidrule(lr){7-11} \cmidrule(lr){12-16} \cmidrule(lr){17-20}
 & {Ori} & \sighl{LDM*} & \sighl{LF++*} & \sighl{LP} & {Ours} & {Ori} & \sighl{LDM*} & \sighl{LF++*}& \sighl{LP} & {Ours} & {Ori} & \sighl{LDM*} & \sighl{LF++*} & \sighl{LP} & {Ours} & \sighl{LDM*} & \sighl{LF++*} & \sighl{LP} &{Ours}\\
\midrule
LGAN++ & 0.387 &  0.598 &0.904 & \seccell{0.360} & \bestcell{0.195} & 0.924 & 1.08 & 0.537 & \seccell{0.281} &\bestcell{0.264}& 0.409 & 0.353 & 0.396 & \seccell{0.413} &\bestcell{0.461} & 0.816 & \seccell{0.895} & 0.477&\bestcell{1.767} \\
BLT  & 0.739 &  0.609 & 1.068 & \seccell{0.392} & \bestcell{0.338} & 0.742 & 1.07 & \seccell{0.633} & 0.645 &\bestcell{0.301} & 0.269 &  0.301 & \seccell{0.366}& 0.342 & \bestcell{0.381} & 0.53 & 0.47 & \seccell{0.551} &\bestcell{1.293}\\
LDM  & 0.633 &  0.624 & 0.913 & \seccell{0.415} &\bestcell{0.201} & 1.01 &  1.19 & 0.632 & \seccell{0.563} & \bestcell{0.369} & 0.307 & 0.309 & \seccell{0.373} & 0.371 & \bestcell{0.415} & \seccell{0.97} & 0.66 & 0.837 & \bestcell{1.46}\\
LF++ & 0.742 & 0.542 & 0.978 & \seccell{0.394} &\bestcell{0.208} & 0.467 &  1.09 & \seccell{0.423} & 0.607 & \bestcell{0.257}  & \bestcell{0.371} &  0.157 & \seccell{0.359} & 0.273 & 0.289 & 0.23 & \seccell{0.327} & 0.287 & \bestcell{1.481}\\
\bottomrule
\end{tabular}
\end{adjustbox}
\caption{
\textbf{Quantitative comparison on content-agnostic generation (Magazine dataset).}
``Ori'' denotes the results generated by each corresponding method.
\rvhl{``LDM*'' and ``LF*'' denote the refinement models based on their original generation model, respectively.}
We highlight the \besthint{first} and \sechint{second} best result for each metric and each layout generation method.
Our method successfully achieve better improvements on alignment, overlay, and containment metrics while preserving the similarity between the input layouts and the rectified layouts.
}
\label{tab:magazine_quan}
\end{table*}


Overall, we alternating between these two stages for $T=5$ iterations.
\sighl{As illustrate in~\mbox{\autoref{fig:two_stage_toy}}, both stages are essential for rectifying flaws.
With only stage A, even though alignments are enhanced, overlaps cannot be mitigated effectively.
}
\sighl{In~\mbox{\autoref{fig:opt_process}}, we show two optimization processes for a document and a magazine layout, respectively.}

\end{description}

\section{Experiment}
\subsection{Datasets and Layout Generation Methods.}

\begin{description}[leftmargin=0pt]
\item[Datasets]
For the content-agnostic layout generation task, we rectify the generated layouts in PubLayNet~\cite{zhong2019publaynet} and Magazine~\cite{zhengsig19} datasets.
We use PubLayNet dataset to demonstrate the ability to resolve undesired overlapping, while we use Magazine dataset to demonstrate the ability to encourage overlap.
We also test our method on content-aware layout generation task using CGL dataset~\cite{cgl2022}.
\sighl{For each dataset, we construct a grid system for each layout in the training data.} 

\item[Generation methods]
\sighl{Across all generation methods, we focus on the ``\textbf{C}ategory$\rightarrow$\textbf{S}ize+\textbf{P}osition'' task,  \ie~the generation method generate the size and position of elements conditioned on the category of each element.}
For content-agnostic generation task, we test our method on the layouts generated by LayoutGAN++ (LGAN++) \cite{kikuchi2021constrained}, LayoutDM (LDM)~\cite{inoue2023layoutdm}, BLT~\cite{kong2022blt}, and LayoutFormer++ (LF++)~\cite{Jiang_2023_CVPR}.
For content-aware generation task, we test our method on the layouts generated by RALF~\cite{horita2024retrievalaugmented}.

\item[Evaluation metrics]
We evaluate the quality of the rectified layouts for all datasets using two metrics: alignment (Align$\downarrow$) and overlap (Ove$\downarrow$) \sighl{\mbox{used in the previous work~\cite{li2020attribute}.}
We provide the detailed definition of them in supplement.}
Additionally, for the magazine dataset, we additionally measure a ``containment'' (Cont$\uparrow$) metric.
Specifically, when an element belongs to ``text-over-image'' or ``headline-over-image'' categories, the containment metric measures the ratio between the intersection area with another element belonging to ``image'' category over its own area.
For CGL dataset, we additionally use ``occlusion'' (Occ$\downarrow$) metric to measure the average saliency value in the overlay region between the background image and the layout elements.
Finally, we measure the IoU similarity between the original layout and rectified layout.
\end{description}

\begin{figure*}[ht]
  \centering
  \includegraphics[width=\linewidth]{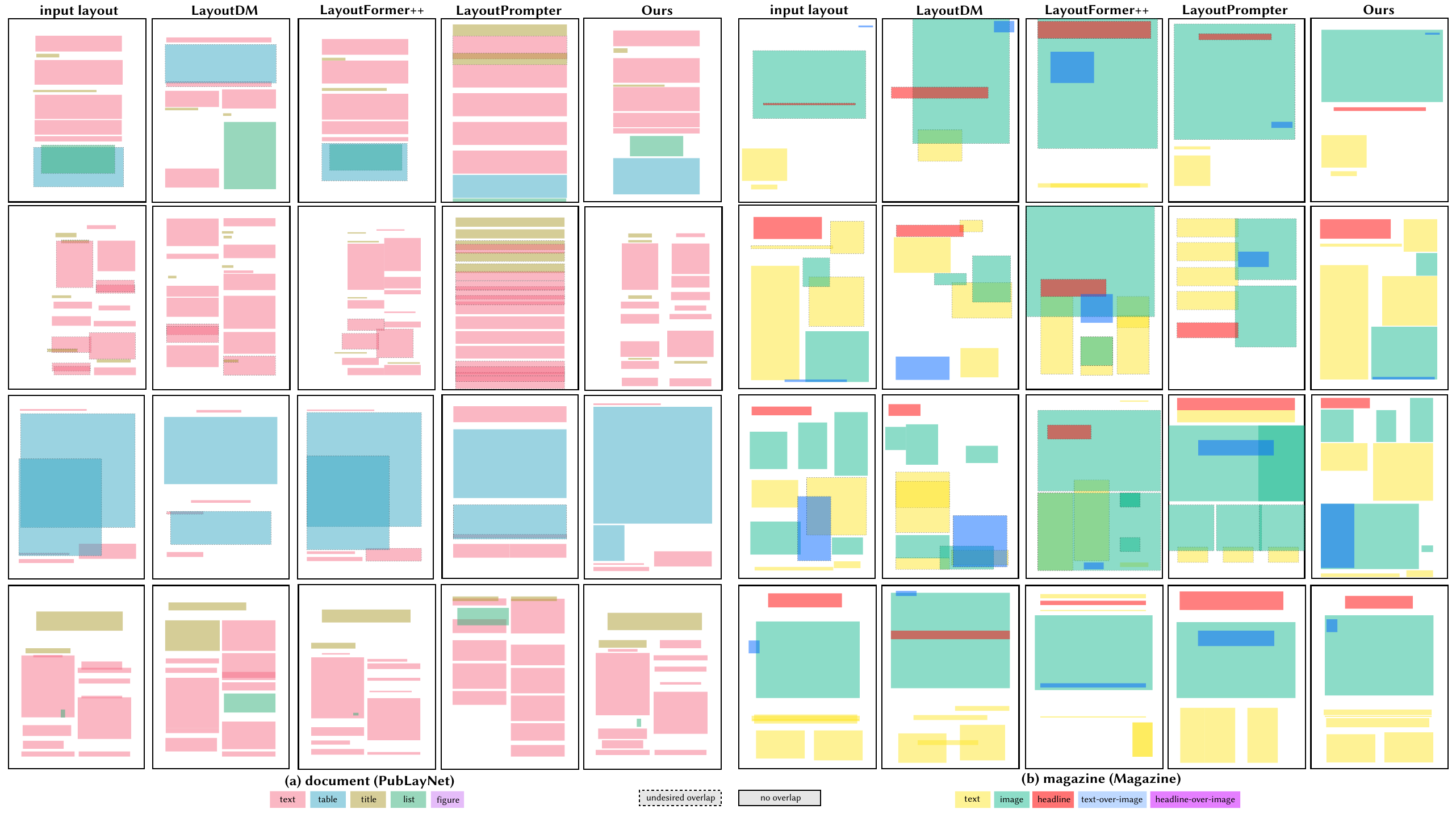}
  \caption{
  \textbf{Qualitative comparison on content-agnostic generation with learning-based and training-free methods.}
  We compared the refined layouts of content-agnostic layout datasets generated by our method with LayoutDM~\cite{inoue2023layoutdm}, LayoutFormer++~\cite{Jiang_2023_CVPR}, and LayoutPrompter~\cite{lin2024layoutprompter}.
  For document layouts, the layouts refined by learning-based methods still have misalignments and unwanted overlaps, while our method generates layouts with better alignment and few overlaps.
  For magazine layouts, learning-based methods struggle to avoid unwanted overlaps and encourage correct containment.
  Moreover, the refined layouts generated by our method closely resemble the original generated layouts.
  }
  \label{fig:learn_compare}
\end{figure*}
\subsection{Content-agnostic Generation.}
We summarize the comparison results of our method on rectifying layouts in PubLayNet and Magazine in~\autoref{tab:publaynet_quan} and~\autoref{tab:magazine_quan}, respectively.
For both datasets, we compared our method with two learning-based layout refinement methods: LDM* and LF++*, as well as one learning-free refinement method: LayoutPrompter (LP).
\rvhl{LDM* and LF* denote the refinement models based on their original generation model.
For LDM*, we retrained the refinement model according to the authors' instructions. 
For LF++*, we used the pretrained refinement models provided by the authors.
}

For the PubLayNet dataset, as shown in~\autoref{tab:publaynet_quan}, our method significantly improves both alignment and overlay metrics.
Even though LF++ already generates layouts with good alignment and few overlays, our method can still rectify the layouts.
When compared to other existing methods, our method obtains the best results in rectifying layouts generated by different automatic generation methods while maintaining the similarity to the original input layout.
In~\autoref{fig:learn_compare}(a), we present four example layouts rectified by three different methods.
The rectified layouts generated by our method have better alignment and less undesired overlays.
\sighl{Our method rectifies a document layout in $0.22$ seconds for seven elements and $0.65$ seconds for 22 elements.}

For the Magazine dataset, as presented in~\autoref{tab:publaynet_quan}, our method enhances all metrics for layouts generated by different learning-based methods.
By minimizing the underlay term (\autoref{eq:containment_term}), our method is able to improve the ``containment'' (Cont$\uparrow$) metric and consistently outperform other existing refinement methods as presented in~\autoref{tab:magazine_quan}.
In~\autoref{fig:learn_compare}(b), we demonstrate that existing methods struggle to resolve flaws in the input layout.
For example, the ``headline'' elements are falsely contained in the rectified layouts generated by LF++ and LDM, and the ``text-over-image'' cannot be fully contained by other ``image'' elements in the LDM refined layouts.
\sighl{It takes $0.71$ seconds on average to rectify a magazine layout with $12$ elements.}

\begin{figure*}[ht]
  \centering
  \includegraphics[width=\linewidth]{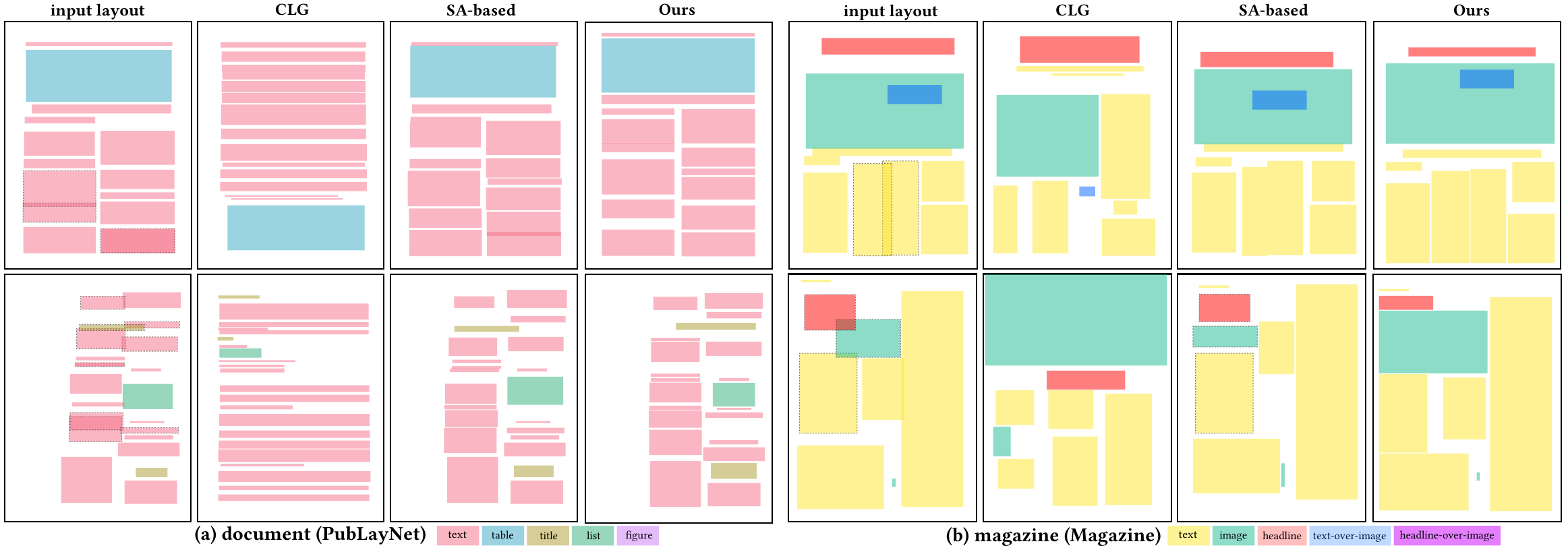}
  \caption{\textbf{Qualitative comparison with optimization-based methods.} 
  The refined layouts generated by our method have better alignment and overlay compared to the results generated by CLG and simulated annealing (SA)-based method.
  Specifically, the layouts rectified by CLG fail to preserve the structure and relations between elements in the input layout.
  \pghl{On the other hand, while the SA-based method rectifies the layout to achieve better alignment and more similar to the input layout, some misalignment and size and aspect ratio discrepancies still persist.
  Moreover, it requires a longer processing time compared to our method.
  }
  }
  \label{fig:opt_compare}
\end{figure*}
\begin{description}[leftmargin=0pt]
\item[Comparison with optimization-based method.]
We further compare the refined layouts generated by our method and \pghl{other  optimization-based methods: CLG}~\cite{kikuchi2021constrained}, LACE~\cite{chen2024towards}, and a simulated annealing (SA)-based method.
The goal of CLG optimization is to obtain a latent vector in the LayoutGAN++ latent space that avoids overlay and improves alignment.
However, since it requires the latent space of LayoutGAN++ generator, we only compare with CLG using the layouts generated by LayoutGAN++.
As shown in~\autoref{tab:clg_comp_quan}, both our method and CLG improve both alignment and overlay.
However, without the underlay term, CLG obtains lower ``containment'' than the original layout ($0.409 \rightarrow 0.291$) because their model tries to avoid all overlay.
More importantly, the refined layouts generated by CLG deviate from the input layout too much according to the similarity metric shown in~\autoref{tab:clg_comp_quan}.
We compute the IoU similarity between the original layout and the refined layout. 
We also demonstrate this issue in~\cref{fig:opt_compare}.
Even though CLG successfully refine the input layout, the relations in the input layout are missing and deviate from the input layout.
\rvhl{Similar to CLG, LACE proposed an optimization strategy using their pretrained layout generator.
In~\mbox{\autoref{tab:clg_comp_quan}}, CLG and LACE show similar performance across different layouts, but both underperform our method on all metrics.
}

\pghl{
On the other hand, we present the rectified layouts by optimizing~\mbox{\autoref{eq:opt_obj}} using a simulated annealing (SA)-based method~\footnote{\url{https://docs.scipy.org/doc/scipy/reference/generated/scipy.optimize.dual_annealing.html}}, as shown in~\mbox{\autoref{tab:clg_comp_quan}} and~\mbox{\autoref{fig:opt_compare}}.
While the rectified results demonstrate reduced misalignment and maintain a structure that is more similar to the input layout, some misalignment issues and discrepancies in element size and aspect ratio remain.
Additionally, the optimization process using the SA-based method takes approximately $30$ seconds to complete, which is significantly longer than our method.
}
\end{description}
\begin{table*}[h]
\small
\centering
\begin{adjustbox}{width=\linewidth, center}
\begin{tabular}{cccccc|ccccc|ccccc|cccc}
\toprule
\multicolumn{1}{c}{} & \multicolumn{5}{c}{\textbf{Align (x100)$\downarrow$}} & \multicolumn{5}{c}{\textbf{Ove$\downarrow$}} & \multicolumn{5}{c}{\textbf{Cont$\uparrow$}} & \multicolumn{4}{c}{\textbf{Sim$\uparrow$}}  \\
\cmidrule(lr){2-6} \cmidrule(lr){7-11} \cmidrule(lr){12-16} \cmidrule(lr){17-20}
\textbf{dataset} & {Ori} & {CLG} & SA & LACE & {Ours} & {Ori} & {CLG}  & SA & LACE & {Ours} & {Ori} & {CLG} & SA & LACE & {Ours} & {CLG}  & SA & LACE & {Ours}\\
\midrule
PubLayNet & 0.18 & 0.149 & 0.129 & 0.115 &\bestcell{0.039} & 0.108 & 0.0018  & 0.0018 & 0.017 & \bestcell{0.0016} &  - & - & - & - & - & 1.299 & 1.734 & 1.413 & \bestcell{1.938}\\
Magazine & 0.924 & 0.61 & 0.508 & 0.801 & \bestcell{0.273} & 0.387 &\bestcell{0.185} & 0.231 & 0.286 & 0.204 & 0.409 &0.291 & 0.326 & 0.387 & \bestcell{0.455} & 0.747 & 1.243 & 0.943 &\bestcell{1.760}\\
\bottomrule
\end{tabular}
\end{adjustbox}
\caption{
\textbf{Quantitative comparison on optimization-based method (CLG~\cite{kikuchi2021constrained}), simulated annealing (SA), and LACE~\cite{chen2024towards}}.
We highlight the \besthint{best} result for each metric and each generation method.
}
\label{tab:clg_comp_quan}
\end{table*}


\begin{figure}[th]
  \centering
    \includegraphics[width=\linewidth]{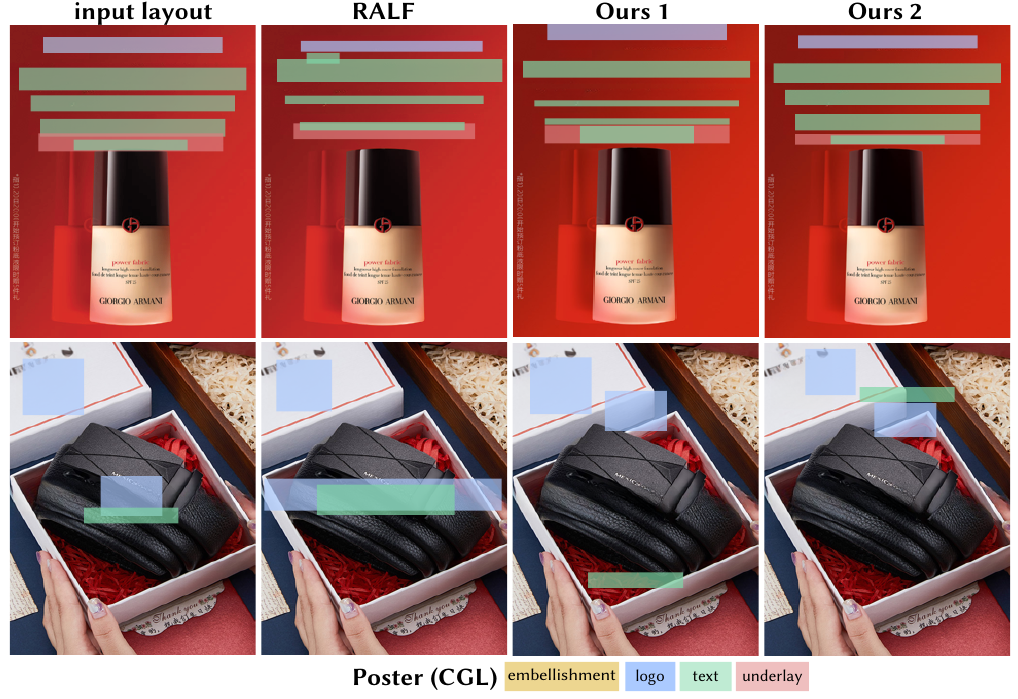}
  \caption{
  \textbf{Qualitative comparison with RALF~\cite{horita2024retrievalaugmented}.}
  We compared the refined layouts of content-aware layout dataset (CGL) generated by our method and RALF.
  Our method resolve the undesired overlapping issue and avoid occlusion over important background image content at the same time, while RALF fails to fix these flaws.
  }
  \label{fig:cont_aware_result}
\end{figure}
\begin{table}
\small
\centering
\begin{adjustbox}{width=\linewidth, center}
\begin{tabular}{c|ccc|ccc|ccc|cc}
\toprule
\multicolumn{1}{c}{} & \multicolumn{3}{c}{\textbf{Ove$\downarrow$}} & \multicolumn{3}{c}{\textbf{Align (x100)$\downarrow$}} & \multicolumn{3}{c}{\textbf{Occ$\downarrow$}} & \multicolumn{2}{c}{\sighl{\textbf{Sim$\uparrow$}}}\\
\cmidrule(lr){2-4} \cmidrule(lr){5-7} \cmidrule(lr){8-10} \cmidrule(lr){11-12}
 & {Ori} & {\sighl{RALF}} &{Ours} & {Ori} & {\sighl{RALF}} & {Ours} & {Ori} & {\sighl{RALF}} & {Ours} & {RALF} & {Ours} \\
\midrule
RALF&  0.006& 0.003 &\bestcell{0.001} & 0.238 & 0.341 & \bestcell{0.133} & 0.126 & 0.125 & \bestcell{0.119} & 1.24 & \bestcell{1.938} \\
\bottomrule
\end{tabular}
\end{adjustbox}
\caption{
\textbf{Quantitative comparison on content-aware generation.}
``Ori'' denotes the layouts generated by RALF.
}
\label{tab:cgl_quan}
\end{table}
\subsection{Content-aware Generation.}
For the content-aware generation, we tested our method on rectifying layouts of CGL.
In~\autoref{tab:cgl_quan}, we compared our method with RALF, a learning-based method.
Our method shows improvement in all metrics, including the occlusion (Occ$\downarrow$) metric, which measures the average saliency value in the overlapping region between the saliency map of the background image and the layout elements.
Moreover, our method outperforms RALF and simultaneously maintains a higher similarity to the input layout.
In~\autoref{fig:teaser}(c) and~\autoref{fig:cont_aware_result}, we show that our method removes undesired overlay and avoids occluding salient image regions.
\sighl{It takes $0.45$ seconds for rectifying a layout with six elements on average.}

\begin{figure}[ht]
  \centering
   \includegraphics[width=\linewidth]{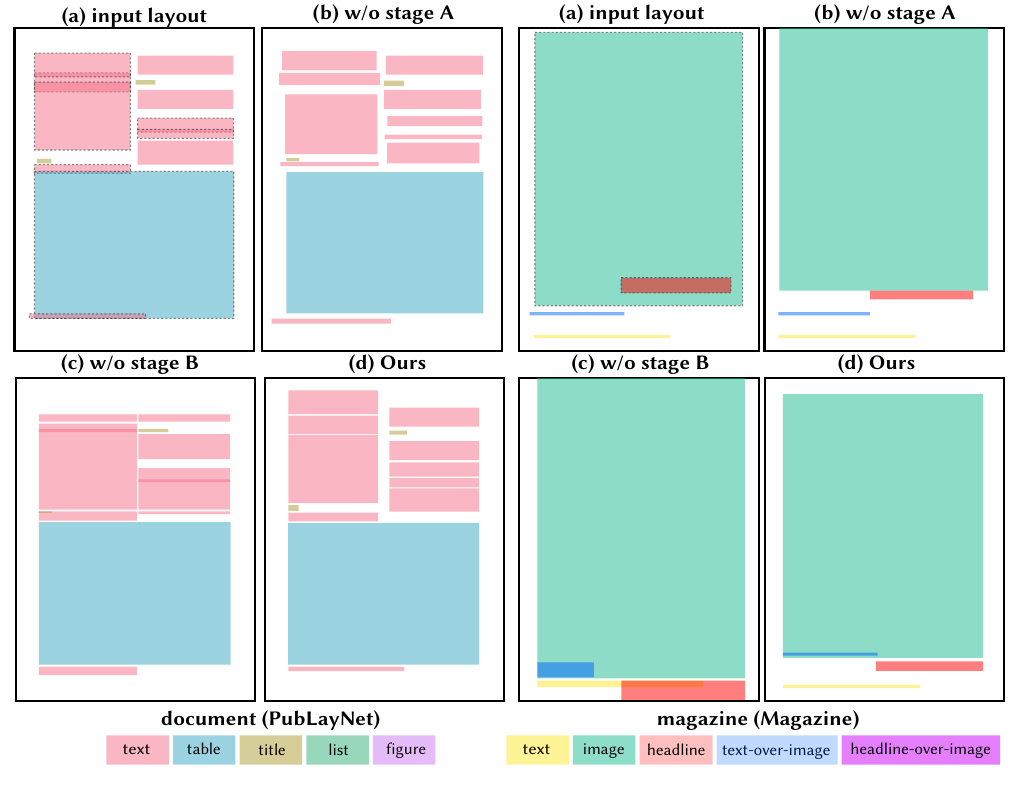}
  \caption{
    \textbf{Ablation study on the two-stage optimization.}
    We compared the refined layouts generated by our method, the ``w/o stage A'', and the ``w/o stage B'' optimization strategy.
    Without stage A, while the unwanted overlaps are mostly removed, the misalignment and containment issues remain unsolved. 
    Conversely, when stage B is excluded, misalignments are effectively corrected, yet unwanted overlaps and containment persist.
    (\protect \blackline: undesired overlaps.)
  }
  \label{fig:ablation_two_stage}
\end{figure}
\subsection{Ablation Study}
\begin{description}[leftmargin=0pt]
\item[The effectiveness of two-stage optimization.]
We compared our two-stage method with two alternative strategies: the ``w/o stage A'' and the ``w/o stage B''.
In the ``w/o stage A'', we optimize~\autoref{eq:opt_obj} directly using a gradient-based optimizer, bypassing the \textit{search-and-snap} strategy.
In contrast, the ``w/o stage B'' relies solely on the discrete \textit{search-and-snap} strategy to optimize~\autoref{eq:opt_obj}.
As shown in~\autoref{fig:ablation_two_stage}, the absence of stage A, the misalignment and containment issues cannot be reduced effectively.
Meanwhile, without stage B, many unwanted overlaps and containment problems persist.
We also conduct quantitative comparisons between our full method and both strategies using the document and magazine dataset.
Overall, our full method achieves better results across all metrics compared to both alternative strategies.
The only exception is ``w/o stage B'' results in slightly better alignment, as focusing exclusively on stage A prioritizes addressing misalignments.
Compared to ``w/o stage A'', our full method obtains $43\%$ better overlap and $41\%$ better alignment and $27\%$ better containment metric.
In comparison to ``w/o stage B'', our full method shows a $68\%$ improvement in overlap and a $23\%$ improvement in containment metric. 
Meanwhile, it results in a $7\%$ worse alignment since optimizing solely with stage A means our focus is primarily on addressing misalignment issues.
Please see supplement for the full quantitative comparison results.

\end{description}

\begin{figure}[t]
  \centering
  \includegraphics[width=\linewidth]{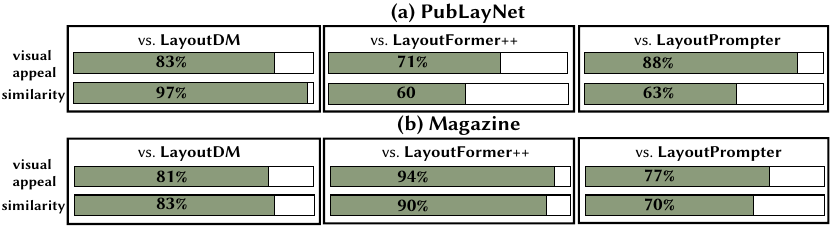}
  \caption{
  \textbf{Results of our user study.}
   The color bar shows the percentage of times that our results are preferred over others.
   Participants consistently prefer refined layouts generated by our method across (a) document and (b) magazine layouts.
  }
  \label{fig:study_result}
\end{figure}
\subsection{User Evaluation}
\sighl{
We conducted a user evaluation on Amazon Mechanical Turk (AMT) following~\mbox{\cite{zhengsig19}}.
We selected $10$ layouts for each dataset and refined them using our method, LDM, LF++, and LP.
Participants evaluated the quality of the generated layouts through pairwise comparisons.
For each layout, we created three comparison pairs, resulting in $30$ pairs for comparison.
In each comparison, participants were asked to select their preferred layout based on \textit{visual appeal} and \textit{similarity to the input layout}.
Each comparison was evaluated by $50$ different workers.
As shown in~\mbox{\autoref{fig:study_result}}, 
participants preferred our method over the others in both criteria.
See supplement for more details.
}
\section{Discussion and Limitations}
\begin{figure}[t]
  \centering
 \includegraphics[width=\linewidth]{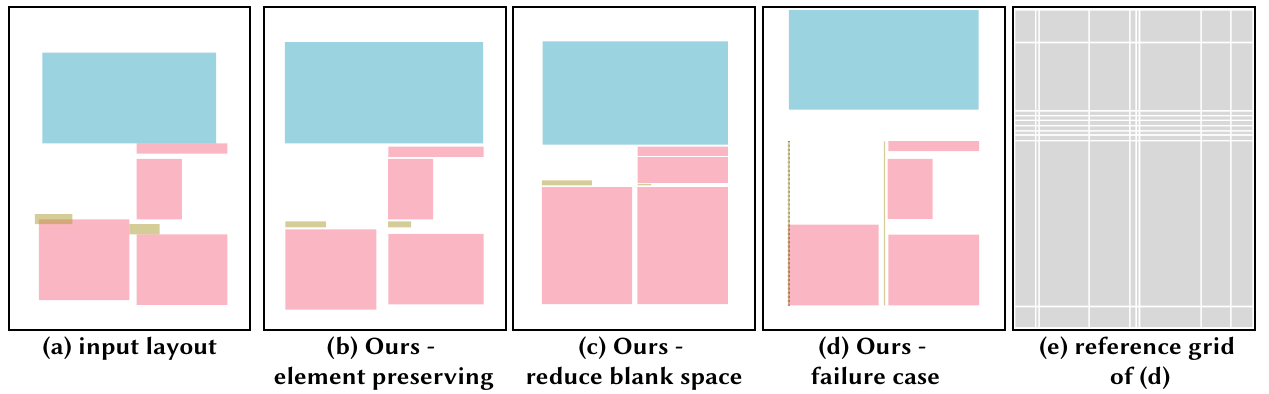}
  \caption{
  When there are large unused spaces in (a) the input layout, our method allows user to decide to refine the layout (b) while preserving element properties or (c) reduce blank spaces by manipulating the weight of the blank space term.
  (d) The result using our method might still contains flaws and deviate from the input layout if (e) the grid system contains redundant and tiny grids which misalign with the structure of the input layout.
  }
  \label{fig:limitation}
\end{figure}

\begin{description}[leftmargin=0pt]
\item[Tradeoff between element preserving and blank space.]
Our method successfully preserves the overall structure, including the alignment relationship, the aspect ratio and area of elements in the original layout.
\sighl{
However, when there are blank spaces in the input layout generated by a learning-based method (\mbox{\autoref{fig:limitation}}(a)), there is a conflict between reducing these blank spaces and preserving the original structure.
To address this situation, our method allows the users to choose whether they want to preserve the element properties (\mbox{\autoref{fig:limitation}}(b)) or reduce the blank space (\mbox{\autoref{fig:limitation}}(c)).
We achieve this by introducing a blank space term $\mathcal{E}_{\text{blank}}$ and optimize it during the \textit{search-and-snap} step.
See supplement for more details about the blank space term.
}
\item[Remaining flaws.]
\rvhl{Our method still cannot rectify all existing flaws. 
This is often due to a conflict between preserving the original layout structure and addressing all flaws.
For example, shrinking an image to avoid overlap may distort its aspect ratio. 
We believe that such trade-offs are best left to users, who can choose to reduce elements or adjust the layout based on their design priorities.
}


\item[Grid mismatch.]
\rvhl{
The quality of grid selection affects rectification performance.
}
Sometimes, the resulting layout's quality can be poor due to an unsuitable reference grid (\autoref{fig:limitation}(d,e)). 
\rvhl{
With thousands of exemplar layouts available,  our retrieval strategy identifies structurally similar grids in most cases. Nonetheless, we recognize this as a limitation and see clear opportunities for improvement.
For example, using a more perceptually informed layout similarity metric like LayoutGMN~\mbox{\cite{patil2021layoutgmn}} could reduce mismatches and enhance overall performance.
Another promising future direction is to generate custom grids tailored to each input layout.
}.

\item[Combination with user interaction]
\rvhl{Our method currently rectifies input layout without user involvement.
In the future, we plan to extend it by incorporating user input, such as adding a penalty for relocating layout elements to main their user-specified positions.
}
\end{description}
\section{Conclusion}
In this paper, we presented \methodName, an optimization-based method for rectifying layouts generated by learning-based methods.
Our approach uses a grid system for structural guidance during optimization, along with a novel box containment cost function.
This function helps prevent unwanted overlap and ensure desired containment \sighl{based on the pre-defined layout-specific criteria}.
We conducted experiments using layouts generated by five different learning-based methods across three public datasets.
Both qualitative and quantitative results demonstrate that \methodName rectifies various flaws with the minimum deviation.
Finally, our method complements existing and future learning-based layout generation methods and does not require any additional training.

\section*{Acknowledgement}
We thank the anonymous reviewers for their valuable feedback.
This work was partially supported by JSPS Grant-in-Aid JP23K16921, Japan and a collaboration with Dentsu Digital.



\bibliographystyle{eg-alpha-doi} 
\bibliography{paper}       


\end{document}